\documentclass[reprint,amsmath,amssymb,prb,floatfix,twocolumn,superscriptaddress]{revtex4-1}
\usepackage[table]{xcolor}
\usepackage{graphicx}% Include figure files
\usepackage{dcolumn}% Align table columns on decimal point
\usepackage{times}
\usepackage{colordvi,epsf}
\usepackage[normalem]{ulem}
\usepackage{soul}

\usepackage{graphicx}
\usepackage{amsmath}
\usepackage{amssymb}
\usepackage{bm}
\usepackage{pgf}
\usepackage{color}
\usepackage{amsfonts}
\usepackage{hyperref}
\usepackage{tcolorbox}
\usepackage[bottom]{footmisc}

\begin{document}

\title{Theoretical investigation on magnetic property of monolayer CrI3 from microscale to macroscale}

\author{Songrui Wei}
\altaffiliation{Songrui Wei\& Dingchen Wang Contributed equally to this work}
\affiliation{Key Laboratory of Optoelectronic Devices and Systems of Ministry of Education and Guangdong Province, College of Optoelectronic Engineering, Shenzhen University, Shenzhen 518060, China}

\author{Dingchen Wang}
\altaffiliation{Songrui Wei\& Dingchen Wang Contributed equally to this work}
\affiliation{MOE Key Laboratory for Nonequilibrium Synthesis and Modulation of Condensed Matter, School of Science, State Key Laboratory for Mechanical Behavior of Materials, Xi’an Jiaotong University, Xi’an 710049, China}

\author{Yadong Wei}
\email{ywei@szu.edu.cn}
\affiliation{Key Laboratory of Optoelectronic Devices and Systems of Ministry of Education and Guangdong Province, College of Optoelectronic Engineering, Shenzhen University, Shenzhen 518060, China}

\author{Han Zhang}
\email{hzhang@szu.edu.cn}
\affiliation{Key Laboratory of Optoelectronic Devices and Systems of Ministry of Education and Guangdong Province, College of Optoelectronic Engineering, Shenzhen University, Shenzhen 518060, China}

\begin{abstract}

\end{abstract}

\maketitle

%\linenumbers

\section*{I. INTRODUCTION}
Magnetic two-dimensional (2D) materials have received tremendous attention recently due to its potential application in spintronics and other magnetism related fields. 
To our knowledge, five kinds of 2D materials with intrinsic magnetism have been synthesized in experiment. They are CrI3\cite{PhysRevX.8.041028,hou2019magnetizing,farooq2019switchable,xu2018interplay}, Cr2Ge2Te6, FePS3, Fe3GeTe2\cite{deng2018gate,fei2018two} and VSe2\cite{bonilla2018strong}. 
Apart from the above intrinsic magnetic 2D materials, many strategies have also been proposed to induce magnetism in normal 2D materials such as atomic modification, spin valve and proximity effect\cite{Kim_2019,gibertini2019magnetic,zeng2019topological}. 
Various devices have also been designed to fulfill the basic functions of spintronics: inducing spin, manipulating spin and detecting spin.

Among the above 2D materials with intrinsic and induced magnetism, CrI3 is a remarkable milestone because it is the first ferromagnetic 2D material down to one atomic layer thickness with an acceptable and tunable Curie temperature\cite{doi:10.1063/1.5096782}. 
It is synthesized in experiment with the mechanic exfoliation method and the magnetism is verified by many measurements such as magneto-optical Kerr effect, X-ray spectroscopy and magnetic circular dichroism\cite{Thiel973,PhysRevX.8.041028}. 
Due to its typical ferromagnetism in in-plane direction and anti-ferromagnetism between neighboring layers, it has promising application in giant magnetoresistance effect, multiferroic phenomenon, magnetic storage and spin valves.

Accompanying with the rapid experimental progress, many simulation and theoretical works have also been performed\cite{kanamori1959superexchange,ehrenberg1998magnetic,goodenough1958interpretation,anderson1950antiferromagnetism}. 
On one hand, for the application purpose, many methods have been proposed to increase the Curie temperature, strengthen the magnetism and stabilize the structure. 
Some devices for the application of magnetic memory storage and magnetic diode have also been designed based on the giant magnetoresistance effect and the concept of magnetic skyrmions. 
On the other hand, to understand the underlying physics, the origin of the magnetism in layered CrI3 have also been widely investigated\cite{goodenough1955theory,guo2018effects,jang2019microscopic,gudelli2019magnetism,sivadas2018stacking,wu2019physical}. 
Some problems such as whether the spin orientation is Ising type or Heisenberg type; is the XXZ model suitable for describing the anisotropy of CrI3; what is the relationship between Kitaev exchange interaction and single-ion anisotropy in CrI3 have been widely investigated\cite{kim2019giant,Kim_2019}.

It has been imagined that around 2025 the magnetic semiconductors may be applied to commercial use. 
Although much progress has been made, most theoretical researches are still limited to the microscopic properties which ban be obtained directly based on density functional theory (DFT) or tight binding model. 
Only a few works try to obtain the macroscopic magnetic properties based on the microscopic results while the investigated macroscopic properties are limited to Curie temperature and critical exponent\cite{liu2019prediction}. 
For applications, the macroscopic magnetic properties such as domain structure and hysteresis loop are important. 
So, a theoretical description of macroscopic properties of magnetic semiconductors with a valid basis will help much in the application of magnetic semiconductors.

The micromagnetic theory is a mature theory which describes the macroscopic properties of magnetic materials such as domain structure and hysteresis loop\cite{doi:10.1063/1.5096782,PhysRevB.99.104432,deb2019topological}. 
Many software based on the micromagnetic theory have also been developed such as OOMMF, mumax3 and spirit. 
One disadvantage of micromagnetic theory is that it needs the empirical parameters as the input and the empirical parameters usually can only be obtained from experiment. 
This strongly impedes the application of micromagnetic theory. 
In this work, we tried to obtain the empirical parameters from first principle simulations based on DFT. 
We constructed the quantitative relationship between the empirical parameters in micromagnetic theory and the energy terms in DFT. 
In this way, the macroscopic properties can be simulated with a valid basis and the disadvantage of micromagnetic theory is overcome at the same time. 
In this work, the microscopic properties include the band structure, density of state, atomic magnetic moment, charge density distribution and exchange coefficient while the macroscopic properties include the Curie temperature, hysteresis loop, domain structure and easy magnetization direction. 
Most of the results are consistent with the experimental or former theoretical work. 
To our knowledge, there is no such an overall investigation of magnetic properties from microscopic to macroscopic on 2D materials. 
Our work will not only facilitate the application of CrI3 in spintronics but also provide a new research method in this field. 

\section*{II. METHOD}
(1) DFT parameters
Our density functional theory (DFT) calculations are carried out within Perdew-Burke-Ernzerhof (PBE) exchange-correlation functional and projector augmented-wave (PAW) pseudopotentials as implemented in Vienna ab initio simulation package (VASP). The cutoff energy for the plane-wave basis is 450 $eV$ and a Monkhorst-Pack k-point mesh of $15 \times 15\times 1$ is used in the hexagonal Brillouin zone. In each configuration, atoms are fully relaxed by employing conjugate-gradient (CG) method. The total energy and atomic forces are converged to 10-5 $eV$ and 0.01 $eV/\AA$ and a large vacuum spacing of at least 20 Å along out-of-plane direction is used in all the calculations.

(2) Transition from output of DFT to input of micromagnetic model
There are at least three parameters that need to be calculated from the results of DFT to perform the subsequent micromagnetic simulation. They are: (a) Saturated magnetization $M_{sat}$; (b) Anisotropy coefficient $K_{u 1}$; (c) Exchange coefficient $A_{ex}$.
(a) Saturated magnetization $M_{sat}$ 
Firstly, the size of atomic magnetic moment can be obtained directly from the OUTCAR of VASP. Secondly, the saturated magnetization $M_{sat}$ can be calculated as: 
\begin{equation}M_{s a t}=\frac{n \cdot M}{V}\end{equation}
where n is the number of magnetic atoms in a unit cell, M is the size of the atomic magnetic moment, and V is the volume of the unit cell. The unit is transformed properly at the same time.

(b) Anisotropy coefficient $K_{u 1}$
The anisotropy energy term in micromagnetic theory describes the total anisotropy of the system. The anisotropy coefficient describes the density of such an anisotropy. Neglecting the higher order anisotropy, the anisotropy coefficient $K_{u 1}$ can be calculated as:
\begin{equation}K_{u 1}=\frac{\Delta E}{V}\end{equation}
where E is the energy difference between the easy magnetization direction and hard magnetization direction of the whole system, and V is the volume of the whole system. In the CrI3 system, the easy magnetization direction is the out-of-plane direction and the hard magnetization direction is the in-plane direction. In mumax3, the easy magnetization direction must also be chosen.

(c) Exchange coefficient $A_{ex}$
The exchange energy density in micromagnetic theory can be expressed as:
\begin{equation}\varepsilon_{e x c h}=A_{e x}(\nabla \boldsymbol{m})^{2}\end{equation}
where $A_{ex}$ is the exchange coefficient and m is the normalized magnetic moment. Considering the relationship between exchange energy density and the difference between ferromagnetic and antiferromagnetic state, the exchange coefficient $A_{ex}$ can be expressed as:
\begin{equation}A_{e x}=\frac{E_{F M}-E_{A F M}}{4 V r^{2}}\end{equation}
where EFM and EAFM is the energy of the system at ferromagnetic and antiferromagnetic state respectively. V is the volume of the whole system and r is the distance between Cr atoms.

(3) The transition from hexagonal structure to rectangular structure
\begin{figure*}[htb]
	\begin{center}
		\includegraphics[width = \columnwidth]{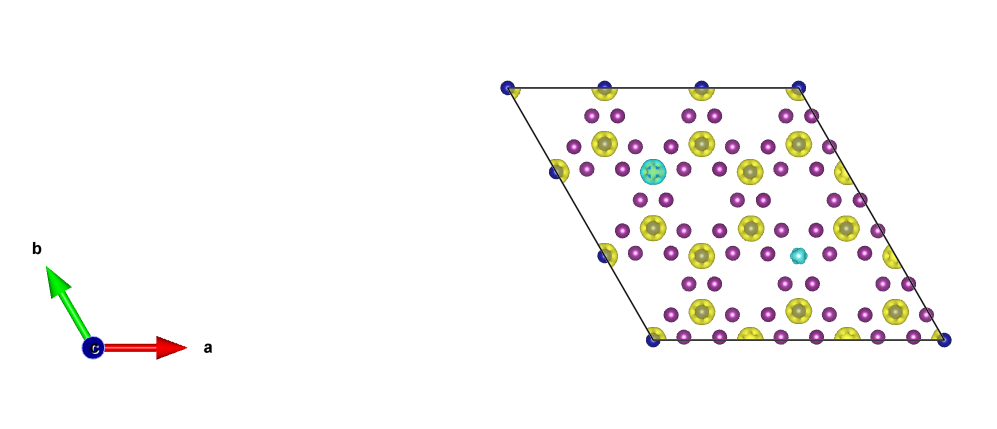}
		\caption{}
		\label{fig:1}
	\end{center}	 
\end{figure*}
In mumax3 and many other micromagnetic software, the basic magnetic moments can only be arranged in rectangular structure. But the structure of many magnetic 2D materials are hexagonal structure. Here we provide a method to transform the hexagonal structure to rectangular structure as shown in \autoref{fig:1}b. We set a 10*10 square of unit cell of CrI3 as the basic magnetic moment of the micromagnetic model and the basic magnetic moments are arranged in rectangular structure. It is worth noting that after the transformation, the above calculated empirical parameters should not be scaled accordingly because all the energy terms in mumax3 are energy densities. 

(4) The micromagnetic simulation

In this section, we will simulate two different sizes of magnetic structure: 2 $\mu m$ $\times$ 2 $\mu m$ and 240 $nm$ $\times$ 240 $nm$. For each size, we will firstly simulate the domain structure and spontaneous magnetization changing with temperature. Then, for certain typical temperature, we will simulate the hysteresis loop and the evolution of domain structure with external magnetic field.

\section*{III. RESULT \& DISCUSSION}
(1)	The first principle calculation part

The magnetic moment, volume of primitive cell and energy terms of different magnetic states of monolayer CrI3 are shown in \autoref{table:1}. In this work, the primitive cell includes 18 Cr atoms. The calculated magnetic moment of Cr atom is in good agreement with its spin quantum number S=3. The volume used here is after the structure relaxation. The energies of different magnetic states are about the primitive cell which includes 18 Cr atoms. These results are consistent with precious theoretical work. \autoref{fig:1}B is the spin density in real space. It is clear that the magnetism is originated from the Cr atoms and the direction of the magnetic moment is in out-of-plane direction.
\begin{table}[]
	\label{table:1}
	\begin{tabular}{llllll}
		\hline
		\multicolumn{1}{|l|}{}     & \multicolumn{1}{c|}{\begin{tabular}[c]{@{}c@{}}Magnetic \\ moment \\ of Cr \\ atom (${\mu}_B$)\end{tabular}} & \multicolumn{1}{c|}{\begin{tabular}[c]{@{}c@{}}Volume of \\ primitive \\ cell (${\AA}^3$)\end{tabular}} & \multicolumn{1}{c|}{\begin{tabular}[c]{@{}c@{}}ferromagnetic \\ state \\ vertical \\ to plane ($eV$)\end{tabular}} & \multicolumn{1}{c|}{\begin{tabular}[c]{@{}c@{}}Ferromagnetic \\ state parallel\\  to plane ($eV$)\end{tabular}} & \multicolumn{1}{c|}{\begin{tabular}[c]{@{}c@{}}Antiferromagnetic\\  state vertical \\ to plane(eV)\end{tabular}} \\ \hline
		\multicolumn{1}{|l|}{CrI3} & \multicolumn{1}{c|}{2.671}                                                                            & \multicolumn{1}{c|}{7335.87}                                                                     & \multicolumn{1}{c|}{-291.31322}                                                                                  & \multicolumn{1}{c|}{-291.29861}                                                                               & \multicolumn{1}{c|}{-291.17477}                                                                                  \\ \hline
		&                                                                                                       &                                                                                                  &                                                                                                                  &                                                                                                               &                                                                                                                  \\
		&                                                                                                       &                                                                                                  &                                                                                                                  &                                                                                                               &                                                                                                                 
	\end{tabular}
\end{table}
(2)	The micromagnetic theory part

\begin{figure*}[htb]
	\begin{center}
		\includegraphics[width = \columnwidth]{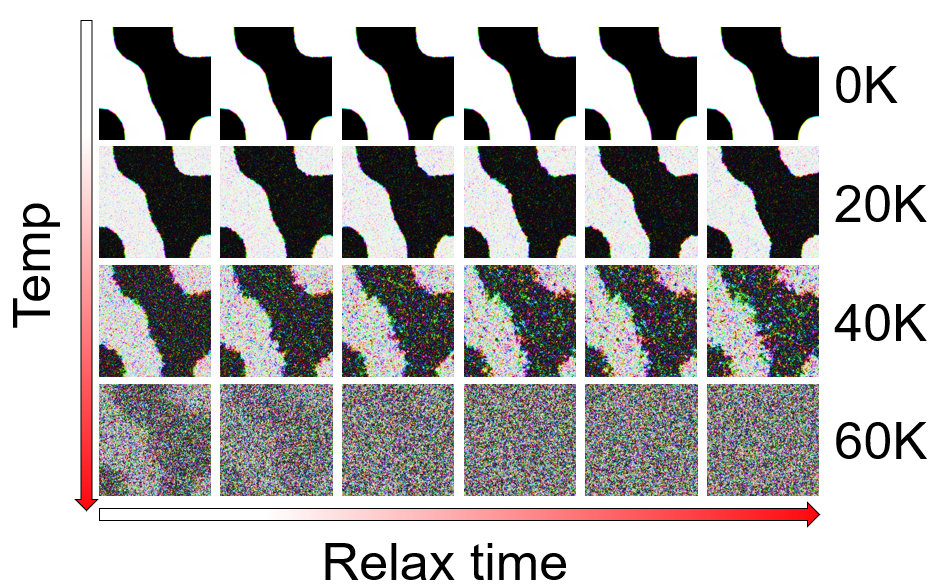}
		\caption{}
		\label{fig:2}
	\end{center}	 
\end{figure*}

\begin{figure*}[htb]
	\begin{center}
		\includegraphics[width = \columnwidth]{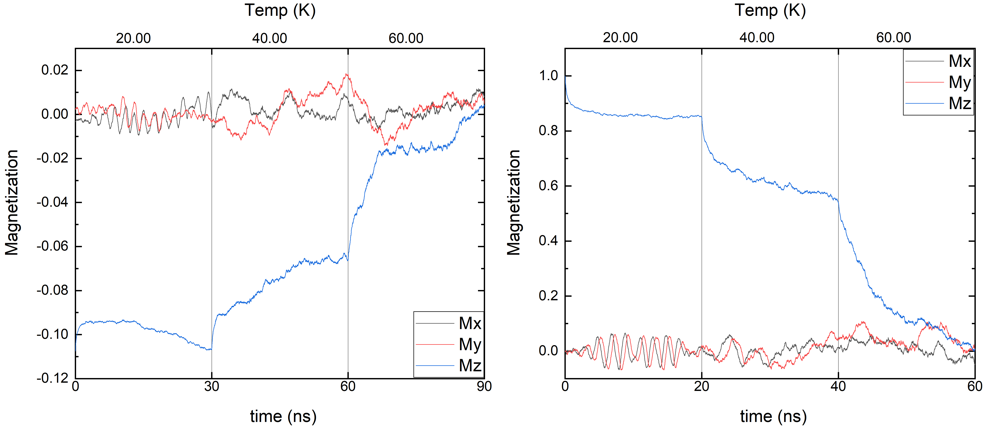}
		\caption{}
		\label{fig:3}
	\end{center}	 
\end{figure*}

The parameters of micromagnetic theory are shown in \autoref{table:2}. They are calculated from the results of first principle calculation. The DM interaction is zero because the structure is symmetrical in vertical direction. Putting the parameters in \autoref{table:2} to mumax3 and the domain structures and spontaneous magnetization in different directions can be calculated. In this work, two different sizes of magnetic systems are calculated. The smaller one is 240 $nm$ $\times$ 240 $nm$ which is in a single domain. The larger one is 2 $\mu m$ $\times$ 2 $\mu m$ which includes more than one domain. On the other hand, the change of domain structure and spontaneous magnetization with temperature is simulated by two steps. In first step, we simulated the domain structures and spontaneous magnetization at the temperature of 0 K, 20 K, 40 K and 60 K, as shown in \autoref{fig:2}. 
The white color and black color represent the “up” and “down” directions vertical to the plane respectively. The other colors mean that there is a component in in-plane direction. At each temperature, the domain structures are fully relaxed. Under such large temperature intervals, the change of domain structure and spontaneous magnetization with temperature or namely the magnetic phase is very obvious. The relaxation process of spontaneous magnetization is also recorded, as shown in \autoref{fig:3}. 

\begin{figure*}[htb]
	\begin{center}
		\includegraphics[width = \columnwidth]{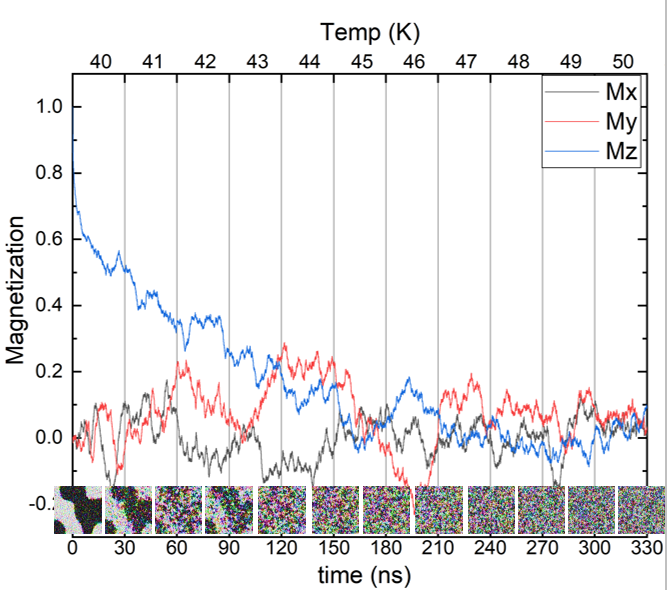}
		\caption{}
		\label{fig:4}
	\end{center}	 
\end{figure*}

The results about domain structure and spontaneous magnetization are consistent with each other. In the second step, we simulated the evolution of domain structure and spontaneous magnetization in the temperature range of 40 K-50 K at the temperature interval of 1 K, as shown in \autoref{fig:4}.
 
By carefully observing the change of domain structure and spontaneous magnetization, the Curie temperature ca be determined as around 43 K. This value is very close to the experimentally measured Curie temperature of monolayer CrI3: 45 K. It is also worth noting that while it is clearer to determine the Curie temperature from the domain structure with a larger sample size (2 $\mu m$ $\times$ 2 $\mu m$), it is clearer to determine the Curie temperature from the spontaneous magnetization with a smaller sample size (240 $nm$ $\times$ 240 $nm$). That is because the domain wall is important for the human eye’s judgement about the transition process while the spontaneous magnetization almost vanishes in a multidomain sample. So, in \autoref{fig:4} the domain structure is about the sample of 2 $\mu m$ $\times$ 2 $\mu m$ while the spontaneous magnetization is about the sample of 240 $nm$ $\times$ 240 $nm$. This is the first reason for why we simulate the samples of these two sizes. 

\begin{table}[]
	\label{table:2}
	\begin{tabular}{llllll}
		\hline
		\multicolumn{1}{|l|}{}     & \multicolumn{1}{c|}{\begin{tabular}[c]{@{}c@{}}Saturated \\ magnetization \\ (A/m)\end{tabular}} & \multicolumn{1}{c|}{\begin{tabular}[c]{@{}c@{}}Anisotropy \\ energy \\ coefficient \\ (J/m3)\end{tabular}} & \multicolumn{1}{c|}{\begin{tabular}[c]{@{}c@{}}Exchange \\ coefficient \\ (J/m)\end{tabular}} & \multicolumn{1}{c|}{\begin{tabular}[c]{@{}c@{}}DM interaction \\ coefficient \\ (J/m)\end{tabular}} & \multicolumn{1}{c|}{\begin{tabular}[c]{@{}c@{}}Easy \\ direction\end{tabular}}            \\ \hline
		\multicolumn{1}{|l|}{CrI3} & \multicolumn{1}{c|}{6.0775E4}                                                                    & \multicolumn{1}{c|}{3.186536E5}                                                                            & \multicolumn{1}{c|}{\begin{tabular}[c]{@{}c@{}}1.06585\\ E-12\end{tabular}}                   & \multicolumn{1}{c|}{0}                                                                              & \multicolumn{1}{c|}{\begin{tabular}[c]{@{}c@{}}Out of \\ plane \\ direction\end{tabular}} \\ \hline
		&                                                                                                  &                                                                                                            &                                                                                               &                                                                                                     &                                                                                           \\
		&                                                                                                  &                                                                                                            &                                                                                               &                                                                                                     &                                                                                          
	\end{tabular}
\end{table}
\begin{figure*}[htb]
	\begin{center}
		\includegraphics[width = \columnwidth]{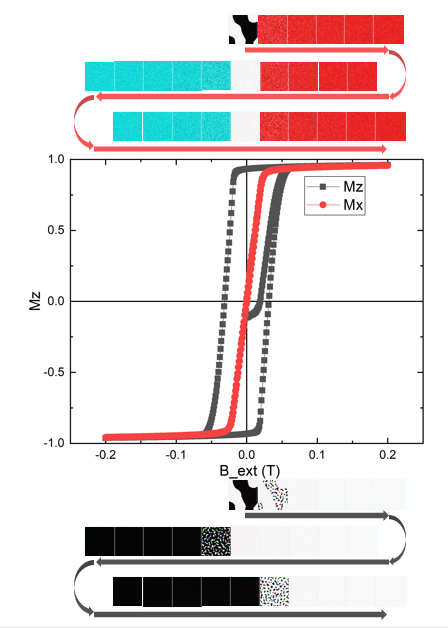}
		\caption{}
		\label{fig:5}
	\end{center}	 
\end{figure*}
For the sample at fixed temperature, the hysteresis loop can be calculated. \autoref{fig:5} shows the hysteresis loop of a 2 $\mu m$ $\times$ 2 $\mu m$ sample at 10 K which is a typical ferromagnetic state. The evolution of domain structures at some magnetic fields are also recorded. When the direction of magnetic field is along the z axis which is vertical to the 2D plane of CrI3 (black line), there is an obvious lag of the response to the external magnetic field and the hysteresis loop is nearly a rectangular. It means that in this case the magnetism is harder. In experiment, the hysteresis loop under a vertical external field is measured with the optical method by many research groups. For the coercivity, the results from experiments and our simulation are consistent with each other. For the magnetization, as the y axis of hysteresis loop measured by an optical method is a percentage which represents the change of magnetization qualitatively, the results from simulation and experiment can not be compared quantitively. When the external field is along the x direction which is parallel to the 2D plane of CrI3 (red line), the area of the hysteresis loop is nearly zero which means that the magnetism is very soft. To our knowledge, the hysteresis loop under a parallel magnetic field has not been measured in experiment yet. The evolution of domain structure with external magnetic field is consistent with the hysteresis loop. 
\begin{figure*}[htb]
	\begin{center}
		\includegraphics[width = \columnwidth]{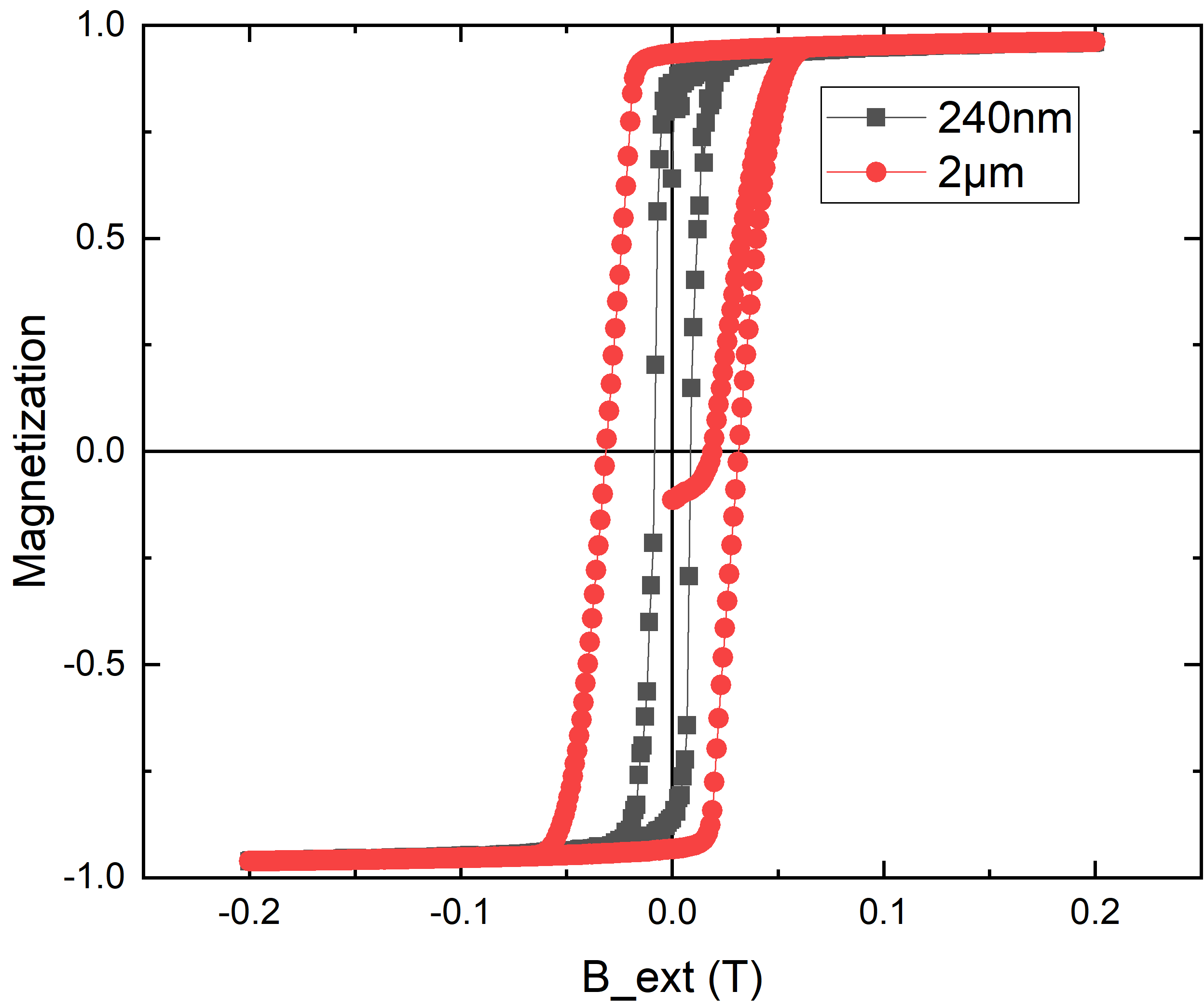}
		\caption{}
		\label{fig:6}
	\end{center}	 
\end{figure*}
There is an obvious lag of evolution of domain structure to external field under a vertical magnetic field. \autoref{fig:6} compares the hysteresis loop of 2 $\mu m$ $\times$ 2 $\mu m$ sample and 240 $nm$ $\times$ 240 $nm$ sample. We did this simulation because in experiment, it is found that the coercivity of monolayer CrI3 measured by optical method may range from 50 to 200 mT. This difference in coercivity is ascribed to the different sample size which may contain one single domain or multi domains. Our result supports this conjecture. 

%
%\bibliography{ref.bib}

\end{document}